# (The) Future (of) Synchrotrons for Particle Therapy

*J. Flanz*
Massachusetts General Hospital, Boston, MA, USA; Harvard Medical School, Cambridge, MA, USA

**Abstract**
The field of particle therapy is quickly growing and yet its more widespread adoption is limited by size, cost, and the need for adaptation to more conformal treatment techniques. In order to realize the benefits of this modality the equipment used to generate and deliver the beam is evolving. The accelerator is one of the key components of this equipment, and its future will be dictated by its ability to accommodate clinical requirements. This lecture is intended to provide an introduction to these requirements and identify how synchrotrons are designed to deliver the desired beams, as well as what limitations exist, and expectations for the future of synchrotrons.

## 1    Introduction

This lecture is intended to offer a perspective on the subject of synchrotrons for particle therapy, about where the field is now and where it needs to go. These questions are answered by reference to an analysis of the requirements of particle therapy.

We start with our present view of the near future. The following are some of the recent themes that have been driving the development of particle therapy.

- *Beam scanning* ('pencil' or 'crayon', PBS). The method of choice for spreading the beam is beam scanning: more particularly, using magnetic fields to move the beam across the target, thus 'painting' the desired area. The size of the 'brush' is the beam size, which is strongly related to the properties of the largely unperturbed beam emerging from the accelerator (and the subsequent focusing systems). The depth of penetration of the beam is primarily determined by the beam energy.

- *Image-guided radiation therapy* (IGRT). The beam position is determined by the use of imaging technology of some sort. For moving targets, the beam properties may require adjustment by feedback during the motion. With protons, in particular, it is possible to image anatomy and directly determine the effective stopping power along the path to the target. Proton radiography and tomography depend upon the ability of the beam to penetrate the patient, and thus require an appropriate beam energy.

- *Adaptive radiotherapy*. Imaging techniques and treatment planning must evolve to a point where a target today that has a different geometry from yesterday (or a minute ago) can be effectively treated. The treatment parameters need to be modified almost on-the-fly. This has implications not only for beam delivery but also for quality assurance.

- *End of range*. Currently, there is some uncertainty in the range of the particles in the patient. This uncertainty results from errors in conversion from X-ray-based imaging and from organ motion or redistribution. Such range information can potentially be obtained more accurately using particle-based imaging or other on-line detection methods, which would then require adjustment of the delivered beam energy during delivery.

– *Ions*. It has been suggested that the treatment of a single tumour could benefit from the use of multiple particles with different values of linear energy transfer, delivered during a single irradiation.

– *Effective cost*. It is a continuing concern that the capital investment is higher for particle facilities than for some other modalities. The basis of that conclusion may be from inappropriate comparisons. In any case, the goal must be to achieve a cost balance in terms of capital investment, patient throughput, and treatment efficacy and accuracy so as to be competitive with other modalities.

Consideration of the above goals of particle therapy, as well as the specific clinical requirements placed on the beam parameters, should be factored into the requirements for the accelerator.

## 2    Flow of requirements

In any discussion of the future of synchrotrons, the above goals should be kept in mind. As part of this, one must clearly define the clinical beam requirements and determine which ones are related to the accelerator design. For some parameters, the characteristics of the accelerator are critical to the beam delivery process, and for other parameters they are almost irrelevant. One must design the accelerator to achieve all the desired clinical goals, not (as in the past) take an existing accelerator and figure out how to apply it to some clinical goals.

The key goals of radiotherapy are:

– to deliver the required dose;
– to deliver that dose with a prescribed dose distribution;
– to deliver that dose in the right place.

The beam delivery system, which is in between the accelerator and the patient, will play a role in how the safe delivery of clinical beam parameters are related to the accelerator parameters. As an example of this, Table 1 shows a few possible parameters and the flow of values from the clinical values to the beam parameters and then to the accelerator parameters that are involved for the case of a beam scanning delivery system.

**Table 1:** Sample of flow from clinical values to accelerator parameters

| Clinical **parameter** | Sample **clinical value** | Beam **parameter** | Accelerator **parameter** |
|---|---|---|---|
| Dose rate | 1 Gy/L min | ~$100 \times 10^9$ protons/min | Beam current |
| Range | 32 cm (in water) | 226.2 MeV protons | Beam energy |
| Scanned-beam penumbra | 80% to 20% fall-off = 3.4 mm (in air) | 3 mm sigma ($e^{-1/2}$ for a Gaussian beam) | Beam size, beam emittance |

As implied by Table 1 and the above text, there is a flow from the clinical requirements and safety requirements to the accelerator requirements. This is depicted in Fig. 1. Starting from any position in the chart other than the top will likely result in compromised treatment parameters.

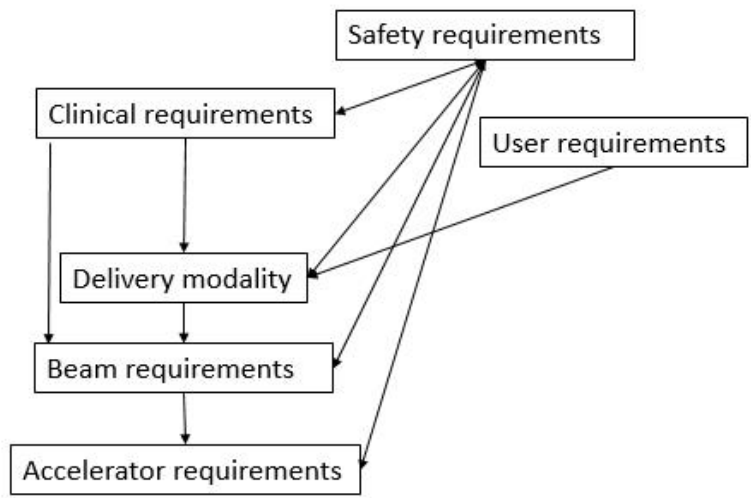

**Fig. 1:** Flow of requirements

## 3 Beam delivery modalities

This lecture is not intended to describe the details of beam delivery modalities for therapy; however, to obtain some basis for the accelerator requirements, it is useful to understand some aspects of this technology.

### 3.1 Beam scattering

A beam-scattering system uses the effects of multiple scattering when a beam passes through a material to spread the beam from the unperturbed 'pencil' to a beam size consistent with the target size. In a double scattering system (a sample of which is shown in Fig. 2), two scatterers with a special profile are used to create a spread-out beam with a uniform transverse distribution as shown in Fig. 3 (lower curve). The beam is also spread longitudinally, to obtain a Spread Out Bragg Peak (SOBP) by selectively degrading the beam energy in the correct proportions to create a flat, spread-out longitudinal distribution as shown in Fig. 3 (upper curve). The properties of the input beam must be tailored to these requirements. However, in this case, because of scattering effects, practically speaking only the initial beam energy is relevant to the beam delivery unless the SOBP requires current modulation to adjust the relative amplitudes of the Bragg peak. The tolerance of the beam position is another factor that needs to be controlled in this beam delivery scenario.

### 3.2 Beam scanning

A beam-scanning system uses magnetic deflection, as shown in Fig. 4, to move the unperturbed beam across the target cross-section, thus spreading out the beam. The unperturbed beam is characterized by a Gaussian profile as in Fig. 5(b), the lower curve. The Gaussian profile is integrated as the beam is moved across the target. Longitudinally the beam profile is that of a Bragg peak as in Fig. 5(a), the upper curve. The energy of the beam is varied, thus adjusting the beam range in such a way as to obtain the desired longitudinal dose distribution.

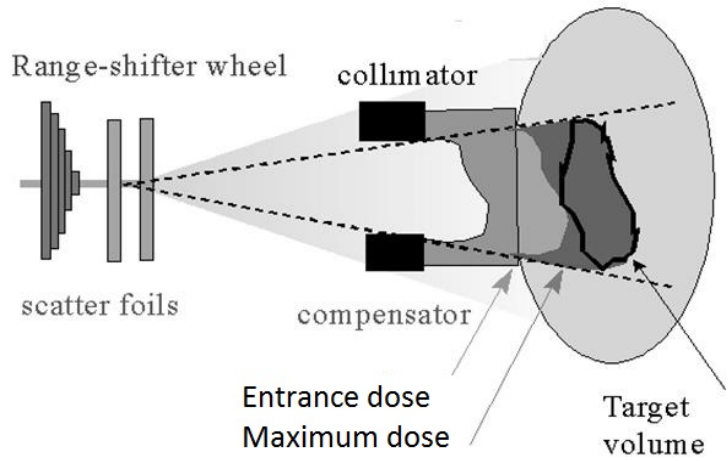

**Fig. 2:** Components of a scattering nozzle

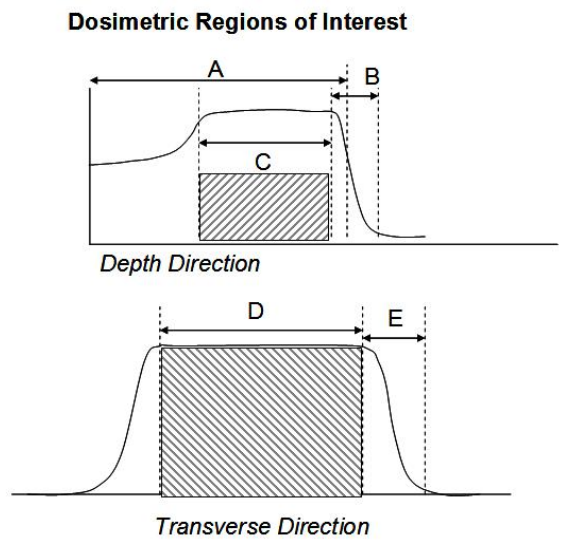

**Fig. 3:** Dosimetric quantities for scattered beams: depth dose (top), transverse dose (bottom)

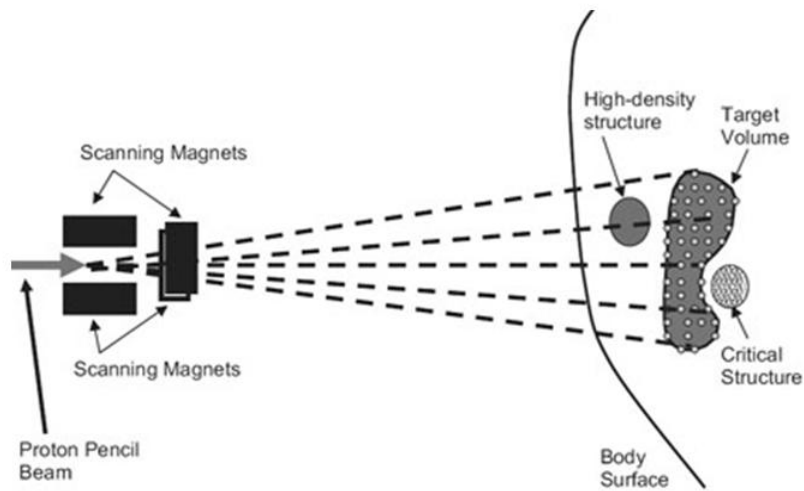

**Fig. 4:** Components of a scanning nozzle

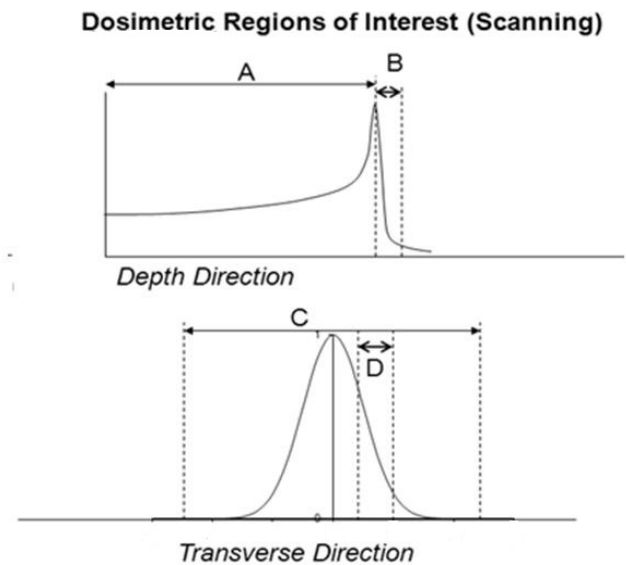

**Fig. 5:** Dosimetric quantities for scanned beams: depth dose (top), transverse dose (bottom)

## 4 Dose and dose rate

The correspondence between an accelerator parameter and a clinical parameter is typically not simple; however, crude estimates may enable an order-of-magnitude value to be determined and used for preliminary accelerator design.

### 4.1 Conversion from beam parameters to clinical parameters

Here is an example of this conversion. We start with the assumption of an accelerator beam of 150 MeV protons and a beam current of 1 nA. This beam is assumed to be incident (via a scanning system) directly (without modification or loss) on a target. Therefore:

- power = joules/second = energy × current
    - e.g., 150 MeV × 1 nA = 0.15 W
- dose = joules/kg ≡ gray (Gy)
    - dose = (power × seconds)/kg
    - e.g., 150 MeV × 1 nA × 60 s = 9 J (for one minute)
- water: 1 kg/1000 cm$^3$ = 1 kg/L
- dose = 9 J/1 kg (in a litre) = 9 Gy
    - 150 MeV, 1 nA = 9 Gy in 1 L in 1 minute
- But not all energy goes into the target (see Bragg peak) ⇒ 3–6 Gy in 1 litre in 1 minute
- 1 nA in 60 s ⇒ 60 × 10$^{-9}$ C ⇒ 3.7 × 10$^{11}$ protons for 3 Gy
- Therefore, for 1 Gy in 1 litre we need ~120 gigaprotons (1.2 × 10$^{11}$)
    - 120 Gp/min ⇒ ~0.3 nA (averaged over a minute, but synchrotrons are cyclic…)

This gives an indication of the number of protons needed to treat a target, depending upon the dose (in Gy) that is prescribed. This number of protons must be extracted from an accelerator in the desired time interval. This lecture is about synchrotrons, so all examples will be relevant to these devices.

## 4.2 Applicability to synchrotrons

A synchrotron is a closed loop of magnetic components in which particles are stored, accelerated, and then extracted (details are discussed in other lectures in this course). During the time the particles are stored and accelerated, they must all live nicely with each other. However, they are all charged, and thus they repel each other. This effect is called the space charge force and is represented in Fig. 6, which indicates the strength of the repulsive/defocusing force across the distribution.

However, as the particles move, they are a medium carrying a current and since parallel currents attract, this attractive force partially cancels the repulsive force, by an amount that depends on the magnitude of the current. As the particles move faster, the deleterious space charge effects are reduced. Thus the worst-case situation occurs during low-energy injection and the number of charged particles that can be stored in the ring depends upon the injection energy of the particles.

Figure 7 shows a graph of the number of protons that can be stored in various medical synchrotrons (the limitation arising from the space charge forces only). Note from the above that 1 Gy/min in a litre $\Rightarrow$ 120 gigaprotons/min $\Rightarrow$ <4 Gp/acceleration cycle (assuming a 2 s cycle), where 4 Gp = $4 \times 10^9$ protons.

Thus a connection between a prescription and an accelerator constraint is obtained.

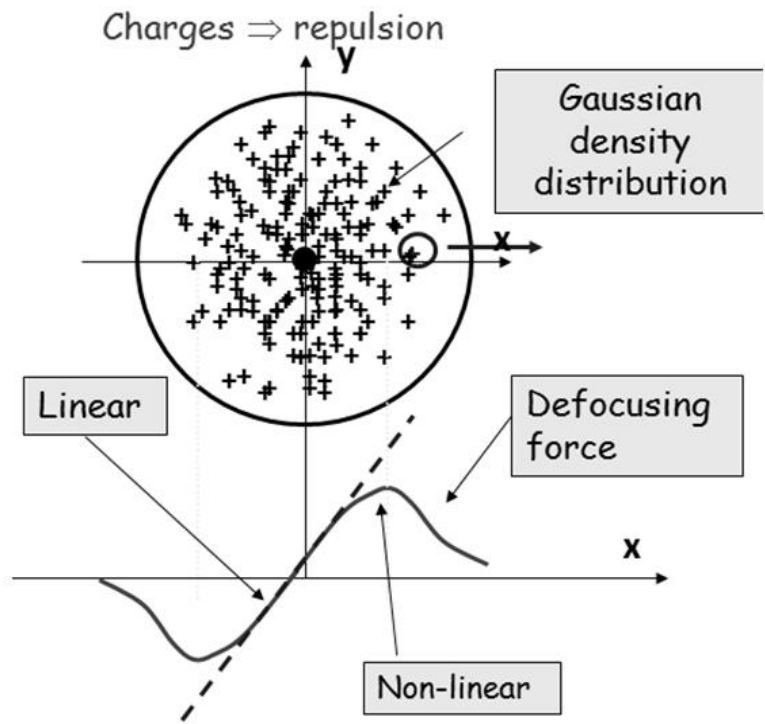

**Fig. 6:** Repulsive forces in a charged beam

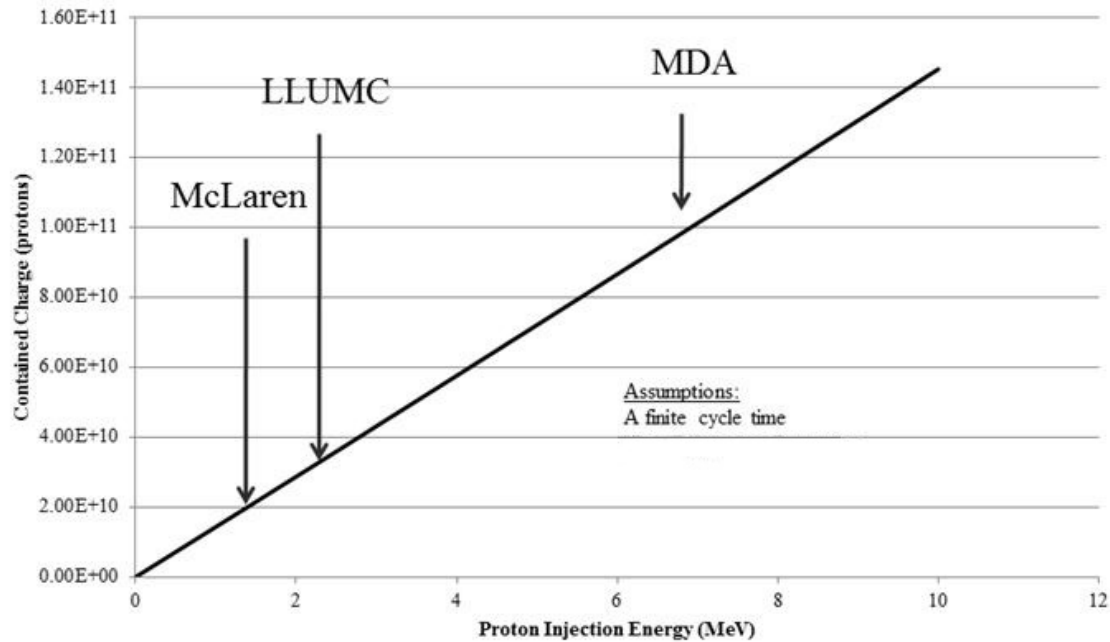

**Fig. 7:** Space charge limits in proton storage rings

### 4.3 Beam current issues

The situation may change depending upon the details of the treatment. A few examples follow.

– The beam-spreading modality plays a significant role. If scattering is used, the beam current incident on the beam delivery system may need to be on the order of nanoamps, whereas for scanning, only tenths of a nanoamp may be required.

– It may be desired to reduce the number of fractions required to deliver the total dose, in which case the dose per fraction would be increased, thus increasing the desired dose rate (so that the treatment time per fraction does not increase).

– Considerations of target motion inside the patient may affect the time constraints on the beam delivery.

– The instrumentation and the beam analysis time will affect the dose rate that can be safely applied.

### 4.4 Tolerances

Depending upon the situation, clinical tolerances can have an impact on the machine performance. Here is one example. Assume that it is desired to deliver 40 gigaprotons to a target (the total dose in a particular field). Clinical tolerance dictates a 2% accuracy in the dose delivery, which results in a tolerance of $8 \times 10^8$ protons. Assume further that this target has transverse dimensions of 10 cm × 10 cm and that the beam spot size used is 5 mm (1 sigma). Thus it will take roughly 20 × 20 beam spots to cover the target. These 400 beam spots will each have to be delivered with a tolerance of $2 \times 10^6$ protons, indicating the level of control required for the beam. Note further that if it takes 100 μs to respond when measuring and reacting to the beam delivery (it could take longer), this tolerance translates to a requirement of not delivering more than $2 \times 10^6$ protons in 100 μs, or a maximum current of 3.2 nA.

## 4.5 Extraction effects

Extracting the beam from a synchrotron is a semi-stochastic process. There are a variety of methods of extracting the beam. One can imagine a pail of water in which water is stored and in which a spigot is inserted into the bottom with a valve to control when, and at what rate, the water is extracted. The water would be expected to come out smoothly, but if it were filled with air bubbles, the result would be different. Figure 8 shows an example of the time dependence of an uncorrected resonant extraction. One class of extraction method is to divide the transverse space inside the synchrotron into a stable and an unstable region. This can be done (in a simplified view) by introducing a non-linear magnetic field, which is weaker than the focusing forces in the ring below a certain radius, but larger at a higher radius (this is a simplified explanation). An example of the kind of time dependence of the extracted beam that can be achieved is shown in the curves in Fig. 9.

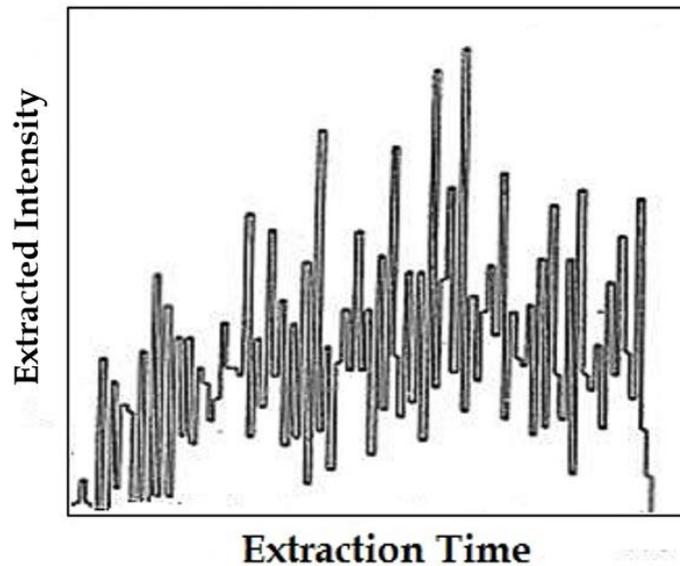
**Fig. 8:** Example of uncorrected time dependence of extracted beam

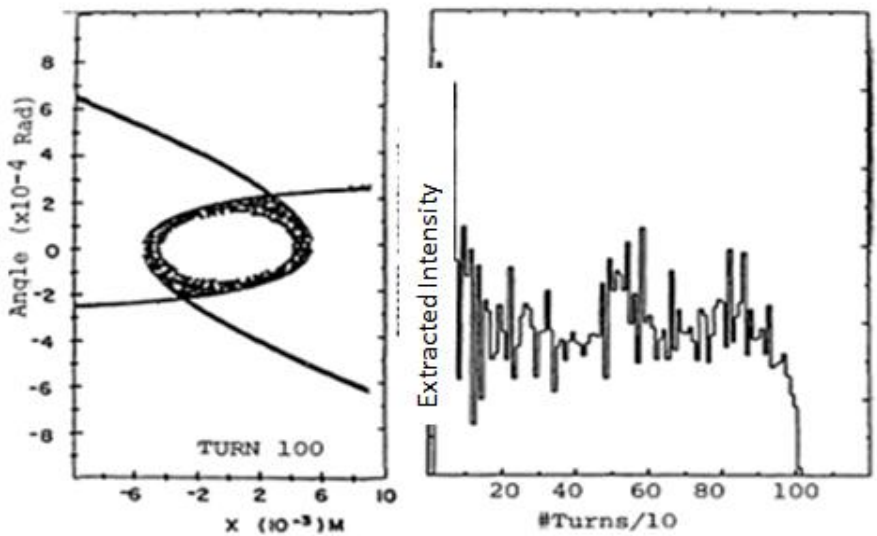

**Fig. 9:** Left: phase space of beam, with separatrix for chromatic extraction. Right: example of time dependence of chromatic extracted beam.

In Fig. 9, the curve on the left shows the phase space of the beam, with the intersecting parabolas representing the phase space boundaries between stable (inside) and unstable (outside) regions. The time dependence of the intensity of a beam extracted from a stretcher ring is shown in Fig. 9 on the right; here, special optics were used in the ring to couple the transverse dimensions to the energy in order to smooth the extraction [1]. Figure 10 shows an example of a scope trace of a beam extracted from a synchrotron used for medical treatment which employs a sophisticated feedback extraction correction system. Issues related to the time it takes to turn off the beam (e.g., when the desired dose is reached) and the range of controllable intensity play a role in determining the appropriateness of an accelerator design.

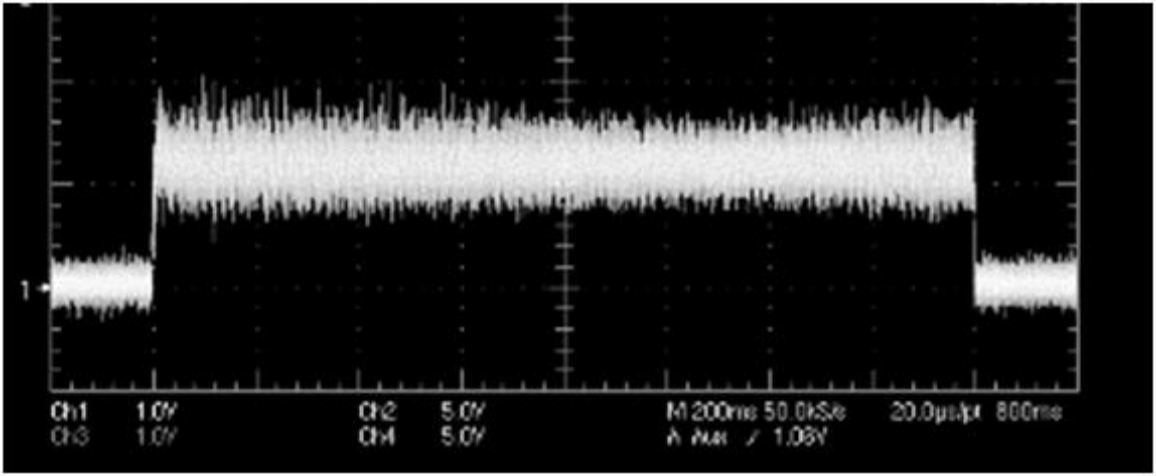

**Fig. 10.** Example of corrected time dependence of RF-excited extracted beam

## 5   Beam size and shape

It is desired to deliver a dose to the target but to allow all surrounding tissue to remain unharmed. The degree of conformity is related to the shape of the beam incident on the target. As noted earlier, the beam may have a Gaussian shape, but it may also have other shapes. A Gaussian shape is particularly well suited for scanned beams since Gaussians can combine well and produce a uniform or otherwise conformal pattern. Figure 11 shows how multiple Gaussians can be combined to form a flat top. The right-hand part of the figure shows that as the Gaussians are separated, the combined dose eventually shows the beam structure.

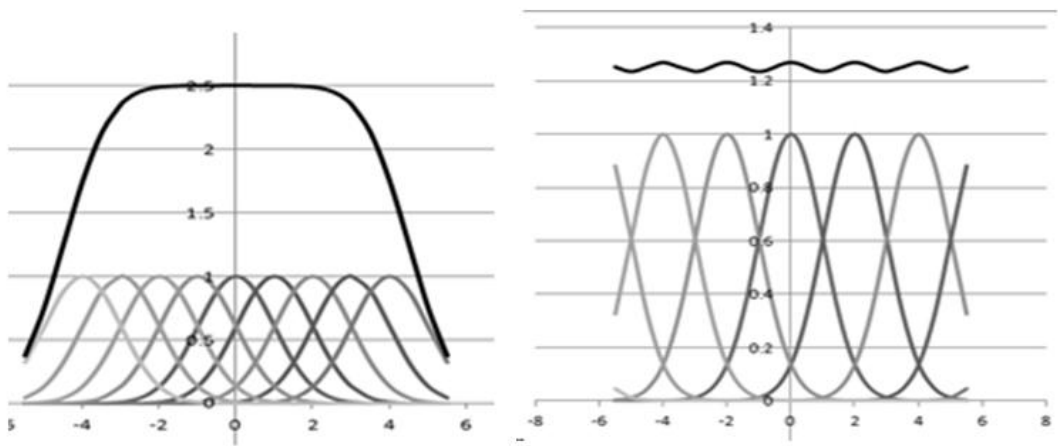

**Fig. 11:** Left: multiple Gaussians spaced such that resultant summed intensity is uniform. Right: multiple Gaussians spaced further apart than optimal, showing the summed intensity structure.

## 5.1 Penumbra

Note that the dose 'falls off' at the sides. The steepest fall-off is at the edge of the Gaussian (without external collimators). The steepness of this fall-off determines how much dose will be deposited into healthy tissue, as shown in Fig. 12. The fall-off region is generally called the penumbra, after the use of this term in scattered-beam physics. If it is desired to ensure that the target dose is within 2.5% of the nominal dose, then the edge of the target must be contained in the top left part of the distribution in the figure (upper box). Suppose, for example, that an important structure occurs in the lower box, to the right; assume, for example, that its edge is 5 mm from the edge of the target. Suppose also that the physician has determined that this critical structure cannot receive more than 50% of the target dose. Using the equation of a Gaussian,

$$e^{-x^2/2\sigma^2},$$

one can calculate that the Gaussian shape cannot have an r.m.s. width (one sigma, $\sigma$) larger than 7 mm.

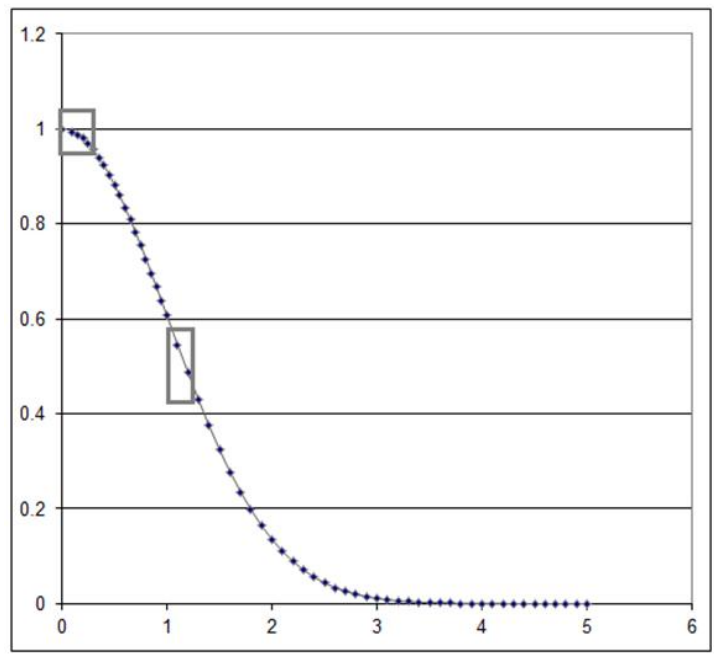

**Fig. 12.** Example of relative intensities of a Gaussian for a given spacing

If the shape of the beam is *not* Gaussian, much of the above discussion becomes invalid and specific calculations are required to determine the adequacy of the beam for treatment. Note that in some synchrotron extraction schemes the beam has a sharp edge in the plane of the extraction (this may occur if there is a septum in the synchrotron).

## 5.2 Effects of emittance

The beam size requirements identified above, coupled with the method by which the beam is delivered (e.g., whether or not it is delivered through a beamline), can place a constraint on the beam emittance. Assuming that it is desired to have a specific beam size at the target and that the final magnetic element is at a fixed distance from the target, a larger-emittance beam will have to be much larger than a smaller-emittance beam at the location of that last magnetic element. If the last magnetic element is on a gantry, then the size and weight of the gantry will be affected by this beam size constraint. Figure 13 shows the power and weight requirements for a gantry dipole located 3 m away from the patient isocentre as a function of the beam size desired. The upper curve is for a beam with an emittance of 25 mm mrad and the lower curve is for a beam with one of 5 mm mrad. Thus one can appreciate the ramifications of larger- versus smaller-emittance beams (independently of how they are achieved).

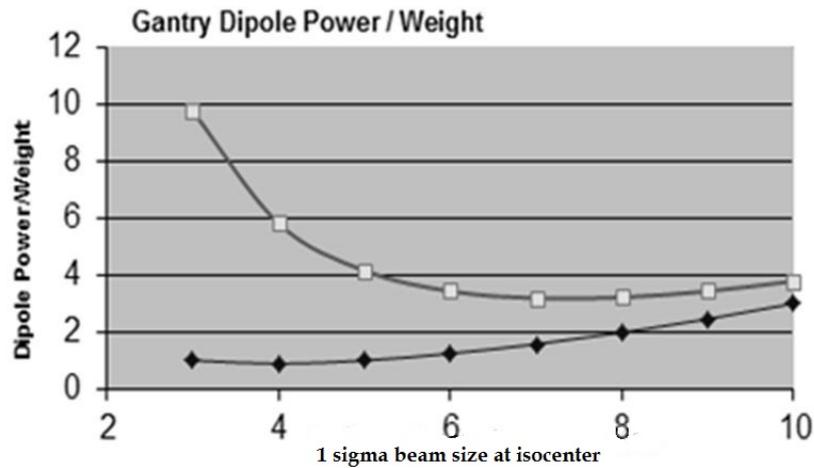

**Fig. 13:** Power and weight of a gantry dipole for a larger-emittance (upper curve) and smaller-emittance (lower curve) beam as a function of the desired beam size at the target.

## 6  Timing

There is a possibility of a strong interaction between the time distribution of the beam extracted from the accelerator and the beam delivery method. In the past few years, beam delivery for particle therapy has evolved into beam scanning as this enables the most conformal dose distribution possible. Therefore we will focus on this aspect of beam timing.

### 6.1  Beam delivery and accelerator timing

There are essentially two styles of scanning beam delivery, one that may be called 'dose driven' and another that may be called 'time driven'. The first integrates the dose at a given location before moving on to the next location. The second assumes that the beam current has the desired stability and uses time as the variable to identify the dose deposited at a given location. If, for example, one moves the beam continuously, the amount of dose deposited along the beam path in any given time interval is determined by the beam current, the scanning speed, or both. If these are not precisely correct, the dose deposited will not be precisely correct. Alternatively, one can deposit a dose at a specific location (spot) and wait until the dose desired there has been delivered (dose-driven method) before moving onto the next location, and therefore the time dependence of the beam (during the time in which the dose is being delivered) is not as relevant, with the following exceptions.

In dose-driven beam delivery, one must stop the beam when the desired dose has been reached. There is, however, a time required for the beam instrumentation to measure and analyse this dose, and a time required for the beam to be turned off. Thus there is a time lag between the time the system decides to turn off the beam and the time the beam is actually turned off. Therefore, either one can anticipate that the beam to be delivered in this time frame will be known and begin the turn-off process earlier, or one can lower the beam current to ensure that the dose that will be delivered in this time frame will not be significant. In addition, in both time-driven and dose-driven continuous scanning, the beam turn-off time will occur while the beam is moving, and one must account for the locations that receive a dose during the turn-off time.

The above considerations highlight the importance of knowing the beam extraction stability and the turn-off time, as discussed above in Section 4.4. The smoothness and controllability of the extracted beam can determine which method of beam scanning can be applied. The smoothness can also determine the anticipated time required to turn off the beam. If the extracted beam current is unstable, one must prepare for the highest beam current and the dose rate must be reduced.

## 6.2  Organ motion

In some cases, the target moves. One hopes to reduce the dose delivered to healthy tissue, and this poses some challenges when the beam delivery has to be done in a time-dependent way. If one could deliver the dose simultaneously to the entire 3D volume of the target (as is almost the case with scattered-beam delivery), one would only need to consider the location of the moving target as a function of time. One could, for example, deliver dose to the entire volume within which the target was moving; then healthy tissue would be irradiated, but the target would receive the desired dose in the shortest possible time. If the beam is gated on only when the target is in the beam path, then the macroscopic timing capability of the accelerator and the time frame of the motion have to be taken into consideration in determining the length of the treatment. Figure 14 shows an example of this situation, which depends upon the time cycle of the accelerator [2]. A large improvement in the efficiency of a synchrotron for beam treatment was achieved with the development of a variable-cycle synchrotron whose injection and extraction can be synchronized with the target motion.

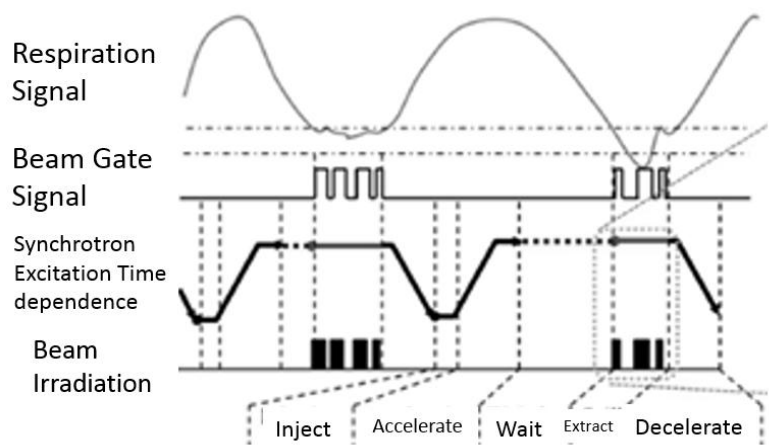

**Fig. 14:** Time dependence of synchrotron operation

An additional effect arises when the beam delivery is done using scanning, which has a 3D position–time dependence. One must consider the timing of the beam delivery, the timing of the target motion, and the timing of the accelerator in order to achieve an appropriate dose distribution in a reasonable time frame.

Some types of beam cycle for synchrotrons include rapid cycling with fast, one-turn extraction or very short pulses of a periodic (e.g., 30 Hz) beam, where each pulse can be at a different energy. Alternatively, one can use slow extraction, pulling out particles from the accelerator as needed, at the beam current needed until they are used up. If there are, for example, $10^{10}$ particles in the synchrotron, then it may take 1 s to use up those particles at the maximum current identified earlier. If more are needed, one has to wait for the time it takes to inject and accelerate another bunch. Also, if a different energy is needed, one must wait for another acceleration cycle, unless one is able to extract at different energy levels during one extraction cycle as shown in Fig. 15 [3]. Note that a breathing cycle can take about 3 s.

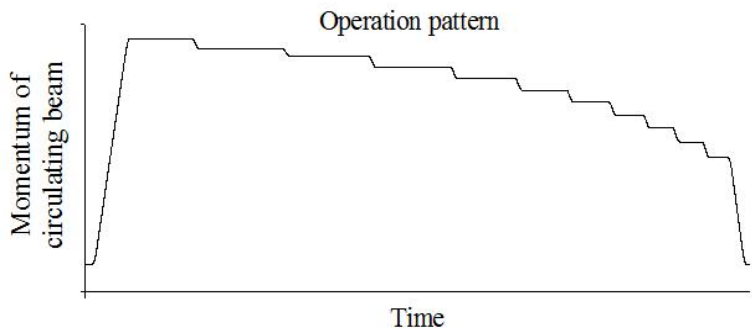

**Fig. 15:** Time dependence of beam energy during extraction in a multienergy extraction process

# 7 Cost

If particle therapy is to become more widely used, the cost of particle therapy systems needs to be reduced. One must reduce both the building and the equipment costs associated with a facility. Today a variety of synchrotrons are being used in medical treatment, some of which are shown in Fig. 16.

Smaller synchrotrons for ions heavier than protons, using superconducting magnets, are being investigated. Proton synchrotrons have already been reduced in size by ProTom and Hitachi (to name two), with diameters on the order of 5 m. Some groups are also attempting to reduce the size and cost of gantry structures. One facility for proton therapy is already being constructed in an existing conventional radiotherapy clinic at Massachusetts General Hospital, with no new building being built for the machine (Fig. 17).

Synchrotrons for heavier ions (e.g., carbon) are much larger, and today resemble a particle physics laboratory accelerator; however, smaller systems utilizing superconducting technologies are under investigation.

**Proton**

| Hitachi 1 | Hitachi 2 | ProTom | Mitsubishi |

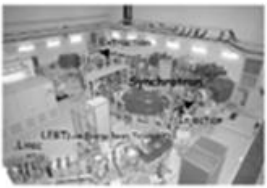 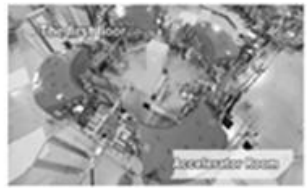 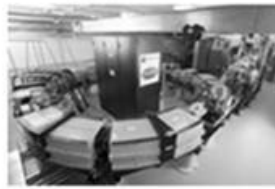 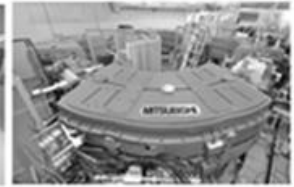

**Heavier Ions**

| Heidelberg | CNAO | Mitsubishi |

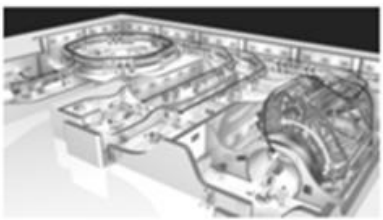 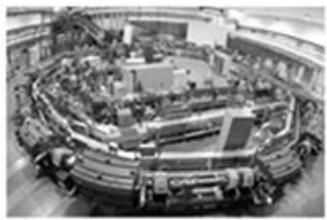 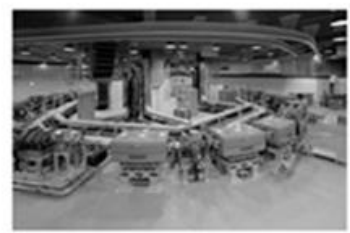

· This does not include all synchrotrons! I apologize for omissions

**Fig. 16:** Some particle therapy synchrotrons currently in operation

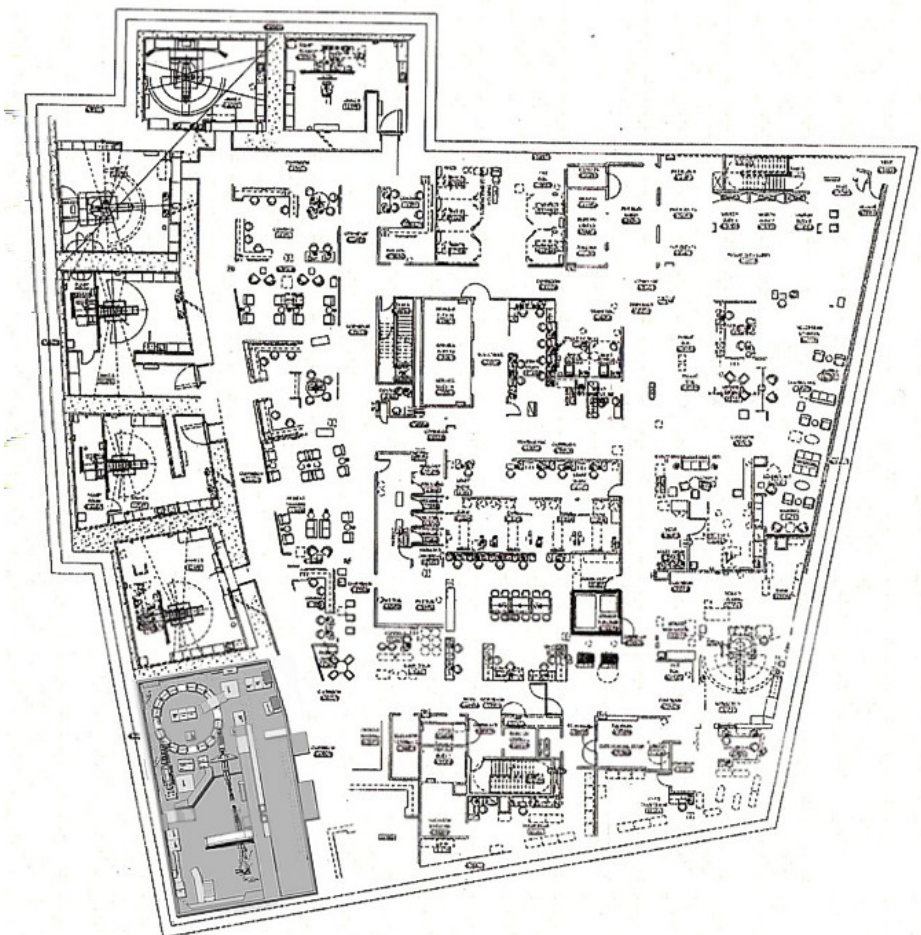

**Fig. 17:** Plan of radiation oncology floor at Massachusetts General Hospital, showing a proton therapy room integrated with linacs (space shaded grey).

## 8 The future of synchrotrons

Many developments of the synchrotron have been necessary in order to develop it into an efficient accelerator for application in medical particle therapy. The design of the synchrotron must correspond to the important treatment parameters. Development work is continuing with the aim of obtaining adequate intensity storage while attempting to minimize the cost of higher-energy injection and smoothly controllable extraction, with a flexible accelerator cycle that can change the beam energy extracted during a single cycle. Thus far, however, all of these capabilities have not been used in any one synchrotron for treatment. For proton synchrotrons, the performance has to continue to improve and the cost has to continue to decrease. There will be increasing demands for faster treatment without compromising accuracy. For heavier particles, the use of superconducting technology can reduce the size, but the cost and the rapidity of change of the magnet excitation will be affected.

There are a number of factors for future consideration (some of which may compete):

- cost:
    - size vs. superconductivity;
    - injector energy;
- intensity:
    - injector energy vs. cost;

- energy:
    - therapeutic energy vs. imaging vs. low (shallow-range) energies;
- energy change speed:
    - effects of superconductivity;
    - beam storage stability;
- turn-off time:
    - instrumentation time;
    - analysis time;
    - extraction control;
- irradiation time:
    - full-volume irradiation in a short time.

The future development of synchrotrons must be directed towards meeting the demands of optimal, safe delivery of particle therapy at a cost that is competitive with conventional radiotherapy systems. It is no longer adequate to identify an accelerator and then ask how it can be used for particle therapy; this has been done. One now has to optimize the delivery of particle therapy, including the beam parameters, timing, size, and cost. Thus far, there is no single standout technology that can accomplish all this, but the technology to achieve operation with the desired parameters does exist and the field is ripe for new insights and developments.